\documentclass{article}
\usepackage{spconf,amsmath,amssymb,epsfig}
\usepackage{tabularx}
\usepackage{booktabs}
\usepackage{enumitem}
\usepackage{float}
\setlist[itemize]{noitemsep, topsep=0.5em, leftmargin=*, labelsep=0.7em}
\usepackage{xcolor}
\usepackage{subcaption}
\usepackage{listings}
\let\OLDthebibliography\thebibliography
\renewcommand\thebibliography[1]{
  \OLDthebibliography{#1}
  \setlength{\parskip}{0pt}
  \setlength{\itemsep}{0pt plus 0.3ex}
}

\usepackage{fancyhdr}
\fancypagestyle{firstpage}{% Page style for first page
  \fancyhf{}% Clear header/footer
  \setlength{\headheight}{15mm}
  \renewcommand{\topmargin}{-20mm}
  \fancyhead[C]{\vspace{5mm} \small IEEE ICME Workshop on Coding for Machines, Niagara Falls, Canada, July 2024\\ © 2024 IEEE. Personal use of this material is permitted. Permission from IEEE must be obtained for all other uses, in any current or future media, including reprinting/republishing this material for advertising or promotional purposes, creating new collective works, for resale or redistribution to servers or lists, or reuse of any copyrighted component of this work in other works.}}
  
\usepackage[margin = 1.9cm, headsep=0.1cm,headheight=0.7cm,]{geometry}
\usepackage{caption}

\expandafter\def\expandafter\normalsize\expandafter{%
    \normalsize
    \setlength\abovedisplayskip{5pt}
    \setlength\belowdisplayskip{7pt}
    \setlength\abovedisplayshortskip{5pt}
    \setlength\belowdisplayshortskip{7pt}
}
\usepackage{xpatch}

\begin{document}\sloppy

% Example definitions.
% --------------------
\def\x{{\mathbf x}}
\def\L{{\cal L}}

% Title.
% ------
\title{Compressive Feature Selection for Remote Visual Multi-Task Inference}
%
% Single address.
% ---------------
\name{Saeed Ranjbar Alvar and Ivan V. Baji\'{c} 
}

%Address and e-mail should NOT be added in the submission paper. They should be present only in the camera ready paper. 
\address{School of Engineering Science, Simon Fraser University, Burnaby, BC, Canada}
%\address{Paper ID 63}

\maketitle

\begin{abstract}
Deep models produce a number of features in each internal layer. A key problem in applications such as feature compression for remote inference is determining how important each feature is for the task(s) performed by the model. The problem is especially challenging in the case of multi-task inference, where the same feature may carry different importance for different tasks. In this paper, we examine how effective is mutual information (MI) between a feature and a model's task output as a measure of the feature's importance for that task. Experiments involving hard selection and soft selection (unequal compression) based on MI are carried out
to compare the MI-based method with alternative approaches. Multi-objective analysis is provided to offer further insight. 
\end{abstract}

\thispagestyle{firstpage}

%============================
% INTRODUCTION
%============================
\vspace{-5pt}
\section{Introduction}
\label{sec:intro}
\vspace{-5pt}
The emergence of high-end mobile/edge devices with Artificial Intelligence (AI) hardware opened new doors for AI applications at the edge~\cite{edgeAI2022review}. However, constraints on the battery and processing power limit the complexity of Deep Neural Networks (DNNs) that such devices can run. One promising solution to this challenge is to divide the DNN workload between the edge device and a more powerful server (e.g., in the cloud), by having the edge device compute features and send them to the cloud for further analysis. This approach is known in the literature under various names, such as collaborative intelligence~\cite{neurosurgeon}, collaborative inference~\cite{shlezinger2022iotm}, computation offloading~\cite{collab_offloading}, split computing~\cite{matsubara2022split}, and semantic communications~\cite{matsubara2022split}. 
Since edge-side DNN features have to be sent to the cloud through a capacity-limited channel, the key challenge becomes how to compress these features without affecting DNN accuracy~\cite{choi_icip,intelligent_sensing}. Increasing interest in these topics has motivated standardization efforts such as MPEG Video Coding for Machines (MPEG VCM)~\cite{MPEG-VCM_DCfE}, MPEG Feature Coding for Machines (MPEG FCM)~\cite{MPEG-FCM_CTTC} and JPEG AI~\cite{jpegai2023mmmag}. 

An important challenge in coding for machines is related to multi-task inference. Multi-task DNNs~\cite{liu-etal-2019-multi,crawshaw2020multi-survey,eva2023cvpr} usually consist of a shared backbone (foundation model) that produces rich features capable of supporting multiple tasks, and multiple heads, each of which performs a certain task. If all tasks are required simultaneously, the solution is to compress all the features. But if only one or a few tasks are required, compressing all features is an overkill; it would be better to find which features are relevant for the particular task(s), and only compress those. This requires the ability to quantify the importance of features for the particular task(s). In this paper, we examine the effectiveness of mutual information (MI)~\cite{Cover} between a feature and a task output as a measure of feature importance, and then describe how it can be used for feature selection. One approach, called \textit{hard selection}, simply retains a set of the most important features and discards others. The other approach, called \textit{soft selection}, compresses some features more and others less, according to their importance.   

The paper is organized as follows. In Section~\ref{sec:related_work}, we review related work on feature selection in machine learning. In Section~\ref{sec:proposed}, we first present our MI estimator for visual multi-task models. We then present our approach to hard and soft feature selection. Experiments are presented in Section~\ref{sec:experiments}, followed by conclusions in Section~\ref{sec:conclusion}.

\vspace{-5pt}
\section{Related work}
\label{sec:related_work}
\vspace{-5pt}
The topic of feature selection has a long history in machine learning~\cite{Feature_selection_JMLR_2003}. In fact, feature selection based on MI is  well-established in  machine learning~\cite{Feature_MI_NCA_2013,MIFS-ND_2014,beraha2019feature}. However, there is little prior work on feature selection in DNNs based on MI, with notable exceptions being~\cite{mint,gradient_MI}, both targeted at filter pruning. Note that feature selection is complementary to feature pruning: selecting a group of features is equivalent to pruning other (non-selected) features. However, filter pruning and feature pruning are not equivalent. For example, consider removing a feature from the output of an \texttt{add} block in a ResNet~\cite{resnet} structure. This is not equivalent to the removal of the corresponding filter in the immediately preceding convolutional layer, because the corresponding channel in the feature tensor is also influenced by the feature carried over through the skip connection. Nonetheless, popular criteria for filter pruning, such as norm-based and proximity-based methods, can also be used for feature pruning (and hence feature selection), by applying the corresponding methodology to features rather than filter kernels. 

Existing approaches to filter pruning can roughly be divided into \textit{norm-based} and \textit{proximity-based} methods. Norm-based methods~\cite{soft_filter,Efficient_ConvNets_ICLR_2017,scale_importance,prun_multi_ss} assume that a large norm is an indicator of the importance of a particular filter in a given network. 
On the other hand, proximity-based approaches~\cite{GM,correlation_filters,spectral_prun} are based on the notion that similar filters may be redundant and can be removed. Feature pruning has followed similar logic. Existing criteria for feature pruning include small variance~\cite{channel_ieee_access}, similarity to other features~\cite{prune_stat_guide}, influence on reconstructing subsequent features~\cite{thinet}, low rank~\cite{HRank}, or attention~\cite{attention_scale}.  

A fundamental problem with the existing feature/filter pruning approaches is that they do not account for the task being performed by the DNN. For example, norm-based methods only consider the characteristics of the feature itself, not what it is used for. Proximity-based methods consider the relationships among features, but again do not account for what they are used for. In a single-task DNN, where all features are used for one task only, this shortcoming may not be obvious, but it is clearly revealed in multi-task DNNs where various features are responsible for different tasks to a varying degree. This is what makes MI an attractive tool for feature selection in multi-task DNNs, because it allows us to connect a feature with a specific task. 

Despite these potential advantages of MI, the scarcity of prior work on MI-based feature selection in DNNs is likely due to the fact that both the intermediate feature space and the output space of a DNN are often high-dimensional, and this is known to be a challenging setup for estimating MI~\cite{Feature_MI_NCA_2013,paninski2003mi,belghazi2021mine,mcallester20aistats}. Practical MI estimators often need to be tailored to the problem at hand. Our work fills this gap and presents an MI estimator suited to visual multi-task DNNs.

\vspace{-5pt}
\section{Proposed methods}
\label{sec:proposed}
\vspace{-5pt}
\subsection{Estimating MI in visual multi-task DNNs}
\label{subsec:estimating_MI}
\begin{figure}[t]
    \centering
    \includegraphics[width=\columnwidth]{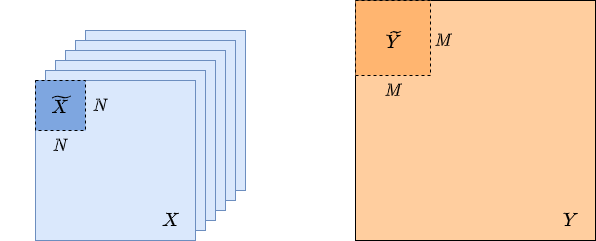}
    \caption{Mutual information between an $N\times N$ feature patch $\widetilde{X}$ and the corresponding $M\times M$ output patch $\widetilde{Y}$ }
    \label{fig:MI_est}
\end{figure}

\vspace{-5pt}
We consider visual multi-task DNNs where the features come in the form of a feature tensor whose channels are feature maps $X$, and the output is an image $Y$, for example, a segmentation map. The high dimensionality of the features and the output makes estimating joint probabilities, and therefore MI as well, challenging due to the ``curse of dimensionality''~\cite{bellman1957dynamic}. To address this challenge, we use the following strategy. First, we divide the output image into $M\times M$ patches (denoted $\widetilde{Y}$) and also divide feature maps into spatially corresponding $N \times N$ patches (denoted $\widetilde{X}$), as illustrated in Fig.~\ref{fig:MI_est}. We then cluster the output patches into $K$ clusters, with cluster index denoted $\widehat{Y}$. This effectively creates an output patch classifier with $K$ classes. If $K$ is reasonably small, one can use MI estimators for DNN classifiers\footnote{We use the binning estimator from~\cite{saxe}.} to estimate $I(\widetilde{X};\widehat{Y})$. Note that both splitting into patches and clustering reduce the MI, so our estimate $I(\widetilde{X};\widehat{Y})$ will be a lower bound on the true $I(X;Y)$.  

To assess the proposed MI estimator, we compare it with the true MI in a few cases where the true MI is known. For jointly Gaussian random variables $X \in \mathbb{R}^n$ and $Y \in \mathbb{R}^m$, with the covariance matrix of %${ \left(\begin{array}{c}X \\Y \end{array}\right)}$ 
$[X,Y]^{\top}$ defined as:
\begin{equation}
\boldsymbol{\Sigma} = \begin{bmatrix}
\boldsymbol{\Sigma}_X & \boldsymbol{\Sigma}_{XY} \\ \boldsymbol{\Sigma}_{YX} & \boldsymbol{\Sigma}_Y 
\end{bmatrix},
\label{eq:multi_gauss_cov}
\end{equation}
where $\boldsymbol{\Sigma}_X$ and $\boldsymbol{\Sigma}_Y$ are covariance matrices of $X$ and $Y$, respectively, and $\boldsymbol{\Sigma}_{XY}$ and $\boldsymbol{\Sigma}_{YX}$ are the corresponding cross-covariances, MI is given by~\cite{MI_formula}:
\begin{equation}
    I(X;Y) = \frac{1}{2} \log \left(\frac{\det(\boldsymbol{\Sigma}_X) \det(\boldsymbol{\Sigma}_Y)}{\det(\boldsymbol{\Sigma})}\right).
\label{eq:true_MI_general}
\end{equation}
%where $\boldsymbol{\Sigma}$ is the $(n+m)\times (n+m)$ covariance matrix for the joint
%variable ($X,Y$). 
%Following the setup in~\cite{MINE}, 
We examine the MI estimates in cases where component-wise correlation between $X$ and $Y$ is given by $\mathsf{corr}(X_i,Y_j) = \delta_{ij} \rho$, where $\rho \in (-1,1)$, and $\delta_{ij}$ is the Kronecker delta.

\begin{figure}[tb!]
\centering
\centering\includegraphics[width=0.4\textwidth]{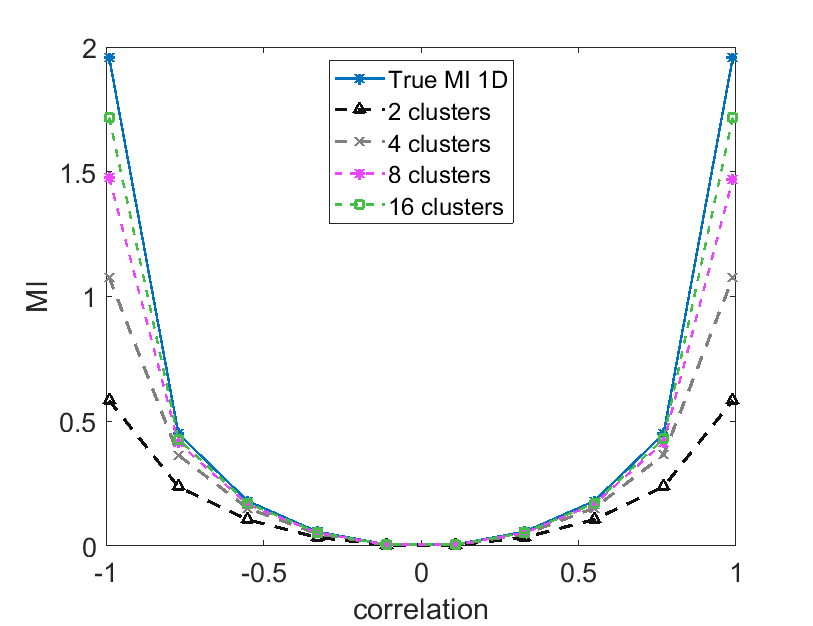}

\centering\includegraphics[width=0.4\textwidth]{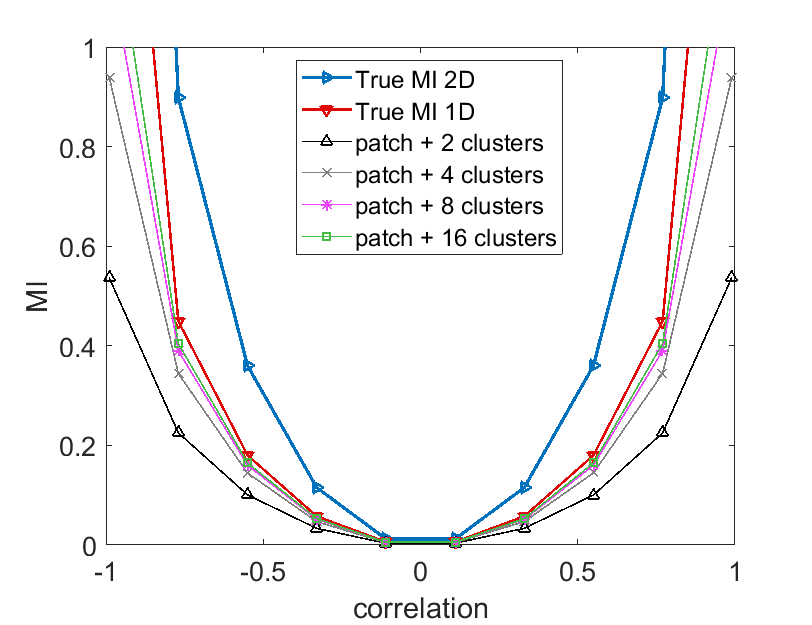}

\caption{True and estimated MI for 1D Gaussian random variables (top) and 2D Gaussian random vectors (bottom).}
\label{fig:MI_estimate}
\end{figure}

First we focus on the scalar case ($m=n=1$), where
\begin{equation}
   \boldsymbol{\Sigma} = \begin{bmatrix}
    1 & \rho \\ 
    \rho & 1 
    \end{bmatrix}
    \label{eq:cov_1D}
\end{equation}
and \eqref{eq:true_MI_general} becomes:
\begin{equation}
    I(X;Y) =  \log \sqrt{\frac{1}{1-\rho^2}}.
\label{eq:true_MI_reduce}
\end{equation}
We examine the effect on the MI estimate of clustering $Y$ into $K$ clusters using K-means clustering~\cite{kmeans}. The true MI and the MI estimates for several values of $K$ obtained from $4\times 10^5$ samples of the mentioned distributions are shown at the top of Fig.~\ref{fig:MI_estimate}. It can be seen that the MI estimates are always below the true MI (i.e., they provide lower bounds), but as the number of clusters increases, the estimate gets closer to the true MI. We repeated the experiments with random state for the K-means clustering 5 times and noticed that the MI estimates are fairly consistent - their variance was $\approx 10^{-8}$.

Patching, as shown in Fig.~\ref{fig:MI_est}, is equivalent to extracting a subspace corresponding to a few coordinates of the original vector. Hence, we examine the effect of both patching and clustering on 2D Gaussian vectors. Specifically, we consider standard isotropic 2D Gaussians $X=[X_1,X_2]^{\top}$ and $Y=[Y_1,Y_2]^{\top}$ ($\boldsymbol{\Sigma}_X = \boldsymbol{\Sigma}_Y=\mathbf{I}$) correlated as above, such that
\begin{equation}
    \boldsymbol{\Sigma}=
    \begin{bmatrix}
    1 & 0 & \rho & 0\\
    0 & 1 & 0 & \rho\\
    \rho & 0 & 1 & 0\\
    0 & \rho & 0 & 1
    \end{bmatrix}.
\end{equation}
In this case, the true MI \eqref{eq:true_MI_general} becomes
\begin{equation}
    I(X;Y) = \log \sqrt{\frac{1}{(1-\rho^2)^2}}.
    \label{eq:true_MI_2D}
\end{equation}
To examine the effect of patching, $X$ and $Y$ are split component-wise, and $X_i$, $Y_i$ are considered samples of patched univariate Gaussian random variables $\widetilde{X}$ and $\widetilde{Y}$. Note that $\widetilde{X}$ and $\widetilde{Y}$ are jointly Gaussian with covariance given by \eqref{eq:cov_1D}, so their MI is given by \eqref{eq:true_MI_reduce}. Since the expression in \eqref{eq:true_MI_reduce} is always less than or equal to \eqref{eq:true_MI_2D}, we see that patching by itself reduces the MI.  Then $\widetilde{Y}$ is clustered into $K$ clusters to obtain $\widehat{Y}$, and $I(\widetilde{X};\widehat{Y})$ is then computed. Fig.~\ref{fig:MI_estimate} bottom illustrates true $I(X;Y)$, the true $I(\widetilde{X};\widetilde{Y}$), and the estimates obtained by clustering $10^6$ samples of $(X,Y)$ and $(\widetilde{X},\widetilde{Y})$.
As before, increasing the number of clusters increases the MI estimate, but because patched univariate variables have a lower true MI than the original 2D variables, the estimates never go above this value. Again, estimates provide a lower bound on the true MI for 2D variables, but the bound is now looser.

In summary, MI estimates obtained by patching and clustering provide lower bounds on the true MI. The bounds are not tight, but they preserve the order -- the higher the true MI, the higher the estimate. Since our feature selection in the remainder of the paper is based on sorting MI estimates, these estimates prove to be useful for our purpose, as will be demonstrated by the experiments.

\vspace{-10pt}

\subsection{Feature selection}
\label{subsec:selection}
Let $X_i$ be the $i$-th feature map and $Y_j$ be the $j$-th task output of a multi-task visual DNN. We define the importance of $X_i$ for $Y_j$ as $I_{i,j} = I(\widetilde{X}_i;\widehat{Y}_j)$,
where $\widetilde{X}_i$ represents $N\times N$ patches from $X_i$ and $\widehat{Y}_j$ represents clustered output patches, as discussed earlier. Note that this definition of feature importance is a function of both the feature $i$ and the task $j$. This is in contrast with established norm-based and proximity-based methods, which are functions only of the feature, but not the task. Further, norm- and proximity-based measures of importance are not scale invariant, because for any scale factor $s\not\in \{0,1\}$, $\lVert s X_i\rVert=|s|\cdot \lVert X_i\rVert \neq \lVert X_i\rVert$ and $ \lVert s X_i - s X_k\rVert=|s|\cdot \lVert X_i-X_k\rVert \neq \lVert X_i-X_k\rVert$. Hence, scaling a feature would seem to change its importance, according to norm- and proximity-based measures. Meanwhile, since scaling by a factor $s\neq 0$ is an invertible operation and MI is invariant to invertible transformations~\cite{Cover}, we have $I(sX_i;Y)=I(X_i;Y)$, showing that our measure of importance is scale-invariant.

\textbf{Hard selection.}
As noted earlier, by \textit{hard selection} we mean retaining only a subset of features. Hence, for a feature tensor with $C$ feature maps, selecting $C'$ features essentially means removing the remaining $C-C'$ features. Suppose we are interested in task $j$. The list of feature importances for task $j$ is $\{I_{1,j},I_{2,j},...,I_{C,j}\}$. Hence, the order of feature importance for task $j$ is
\begin{equation}
   \mathcal{O}_j = (O_{1,j}, O_{2,j}, ..., O_{C,j})=\mathsf{sort}\{I_{1,j},I_{2,j},...,I_{C,j}\}.
   \label{eq:order}
\end{equation}
Hard selection then simply means retaining the $C'$ top-ranked features from $\mathcal{O}_j$.

\begin{figure}[tb]
    \centering
    \includegraphics[width=\columnwidth]{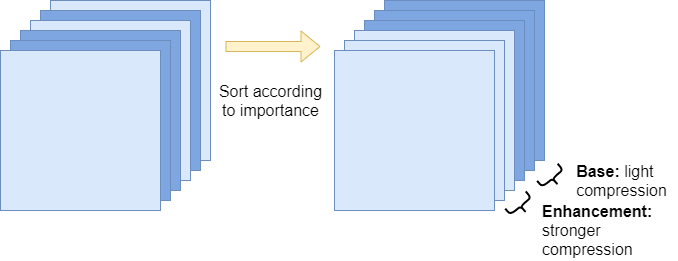}
    \caption{Illustration of soft feature selection.}
    \label{fig:soft_pruning_illustration}
\end{figure}

\textbf{Soft selection.}
In this work, \textit{soft selection} refers to the following process. First, a certain number of features is selected as the most important (or \textit{base}) features, and the remaining features are declared less important (or \textit{enhancement}) features. Base features are quantized to 8-bits, but not compressed any further. Meanwhile, stronger compression is applied to enhancement features, as illustrated in Fig.~\ref{fig:soft_pruning_illustration}. With the notation defined above, if we are interested in task $j$, then the $C'$ top-ranked features from $\mathcal{O}_j$ will be declared base features and the remaining $C-C'$ features will be declared enhancement features, and compressed more heavily. This way, enhancement features are not completely removed as in hard selection, but compressed instead. Moreover, the amount of compression can be controlled, which allows this approach to operate in the range between complete feature removal (hard selection) and full feature preservation. This is a useful property to have when the constraint on the system is specified as the maximum bitrate, rather than the number of features. 

In terms of implementing soft selection, 
any feature compression method can be used for the enhancement features. In this work, the enhancement features are first tiled and then compressed using a High Efficiency Video Coding (HEVC) codec, as in~\cite{choi_icip}. In HEVC, quality of the compressed signal is controlled via the quantization parameter QP, and the experiments will examine the behavior of the model with various values of QP used for the enhancement features.

\begin{figure}[t]
    \centering
    \includegraphics[width=\columnwidth]{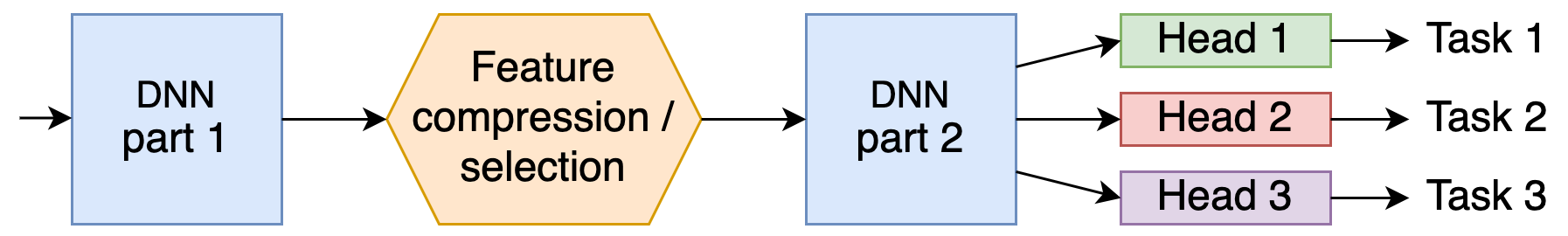}
    \caption{Three-task DNN from~\cite{Saeed_ICASSP_2020} used in our experiments. Heads 1-3 are similar to FC8 networks~\cite{FC}. }
    \label{fig:summary}
\end{figure}

\begin{figure*}[tbh]
\includegraphics[width=5.8cm]{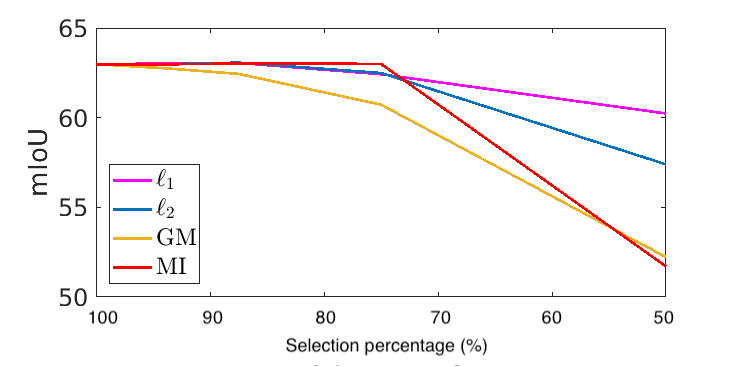}
\includegraphics[width=5.8cm]{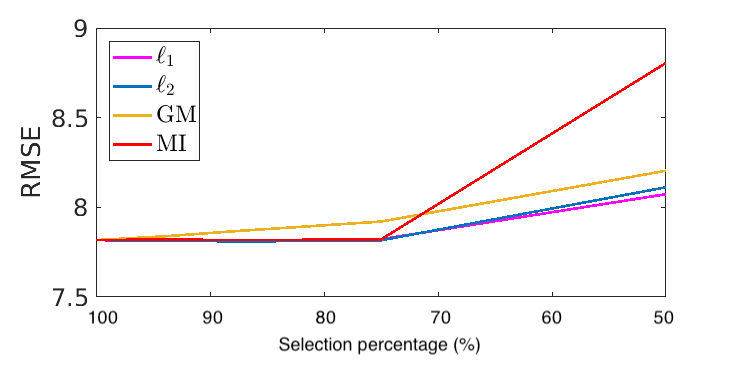}
\includegraphics[width=5.8cm]{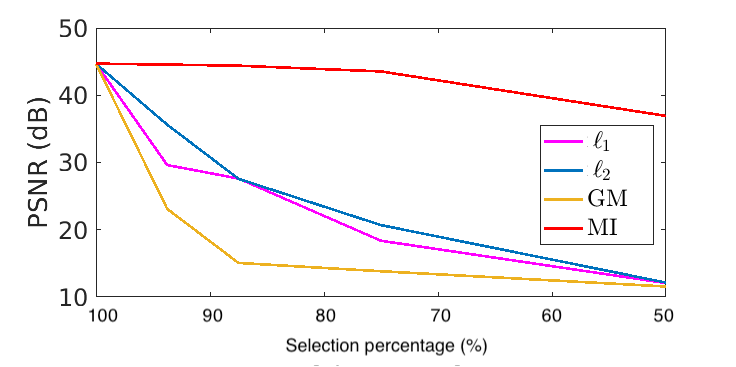}
\caption{Task accuracy vs. hard selection \% for semantic segmentation (left), disparity map estimation (middle) and input reconstruction (right)} 
\label{fig:multi_hard}
\end{figure*}

\begin{figure*}[t]
    \centering
    \includegraphics[width=\linewidth]{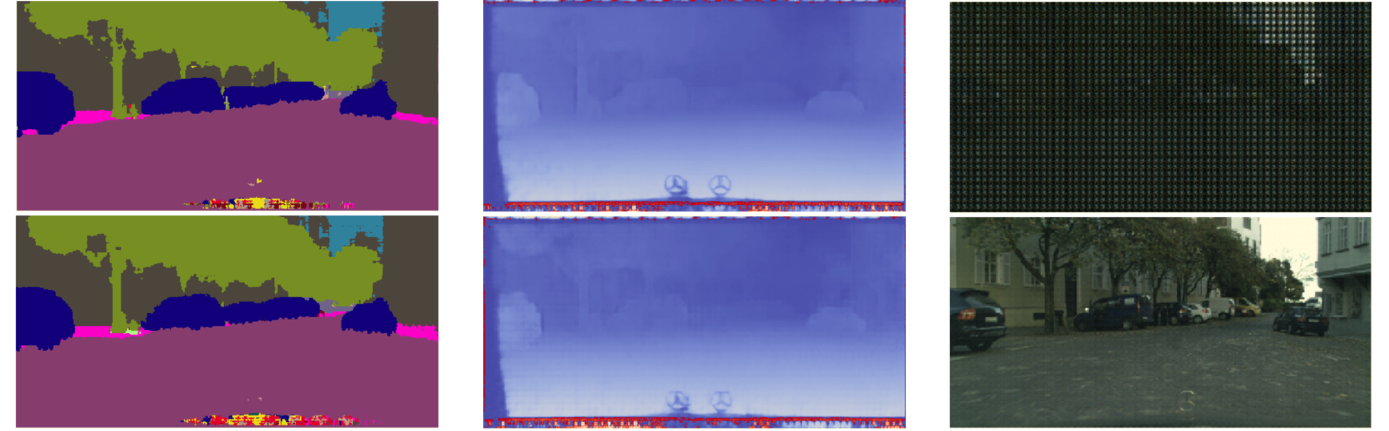}
    \caption{An example of task outputs with $50\%$ hard selection, left to right: semantic segmentation map, disparity map, and reconstructed input. Top: $\ell_2$ norm-based selection. Bottom: MI-based selection.}
    \label{fig:l2_vs_MI}
\vspace{-5pt}
\end{figure*}

\vspace{-5pt}

\section{Experiments}
\label{sec:experiments}
\vspace{-5pt}
Our experiments are based on the multi-task DNN from~\cite{Saeed_ICASSP_2020}, which is illustrated in Fig.~\ref{fig:summary}. This model was trained on Cityscapes~\cite{Cordts2016Cityscapes} to perform three tasks: (1) semantic segmentation, (2) disparity map estimation, and (3) input image reconstruction on Cityscapes. The backbone is similar to the backbone of YOLOv3~\cite{YOLOv3} with 74 layers. The split point is layer 36 of the backbone, which outputs features of size $32\times64\times256$, i.e., $C=256$. The metrics used to assess the performance of the model on the three tasks are as follows: mean Intersection over Union (mIoU)%~\cite{metrics} 
for semantic segmentation, Root Mean Squared error (RMSE) in pixels%~\cite{metrics} 
for disparity maps estimation, and Peak Signal to Noise Ratio (PSNR) in dB for input reconstruction.

\textbf{Hard selection.}
We compare our MI-based feature selection with selection based on $\ell_1$ norm~\cite{Efficient_ConvNets_ICLR_2017}, $\ell_2$ norm~\cite{soft_filter}, and geometric median~\cite{GM}. 
We used $N=8$ and $M=64$ (Fig.~\ref{fig:MI_est}) for feature and output patches, respectively, and further clustered  
output patches into $K=8$ clusters using K-means clustering.   Fig.~\ref{fig:multi_hard} shows the task accuracies at five selection percentages
$\{100, 93.75, 87.5, 75, 50\}\%$ for the benchmarks and the proposed method. 
As seen in Fig.~\ref{fig:multi_hard} (a)-(b), norm-based selection and the proposed MI-based selection provide similar performance up to $75\%$ in semantic segmentation and disparity map estimation (with RMSE, the lower the better), while the GM-based method is slightly worse. At $50\%$ selection on these two tasks, norm-based selection achieves better performance than the other two approaches, but the differences are relatively small in relative terms. However, on the input reconstruction task (Fig.~\ref{fig:multi_hard} right), MI-based selection achieves significantly better performance compared to the other approaches across all selection percentages. This highlights the fact that the norm- and similarity-based importance cannot capture the feature importance for all tasks in a multi-task model; they may work well for some tasks, but not others. Fig.~\ref{fig:l2_vs_MI} shows the task outputs with $\ell_2$-based and MI-based feature selection. The outputs for semantic segmentation and disparity estimation are similar, but input reconstruction with MI-based selection is much better. In the remainder of the experiments, we use $\ell_2$-based selection as a benchmark.

\begin{figure*}[h]
%\begin{tabular}{ccc}
\centering
\includegraphics[width=5.5cm]{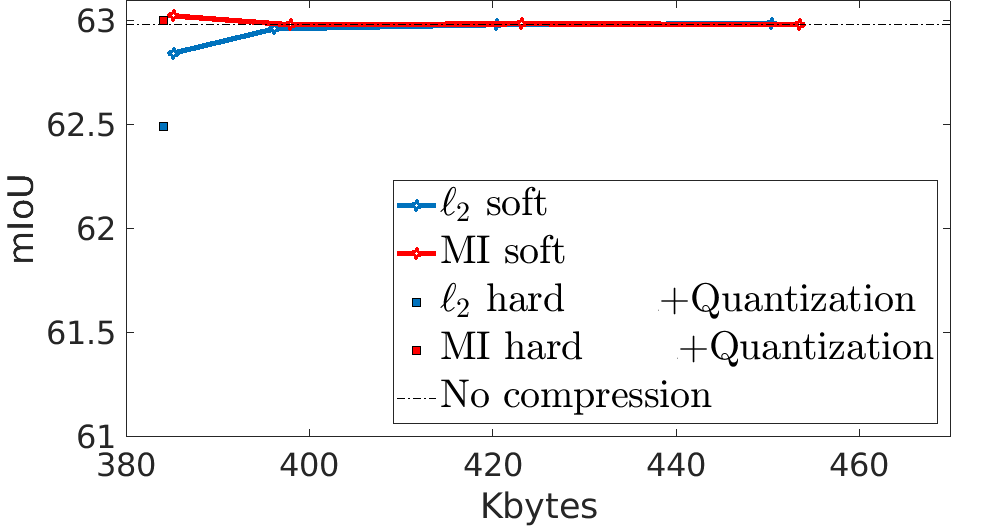}
\includegraphics[width=5.5cm]{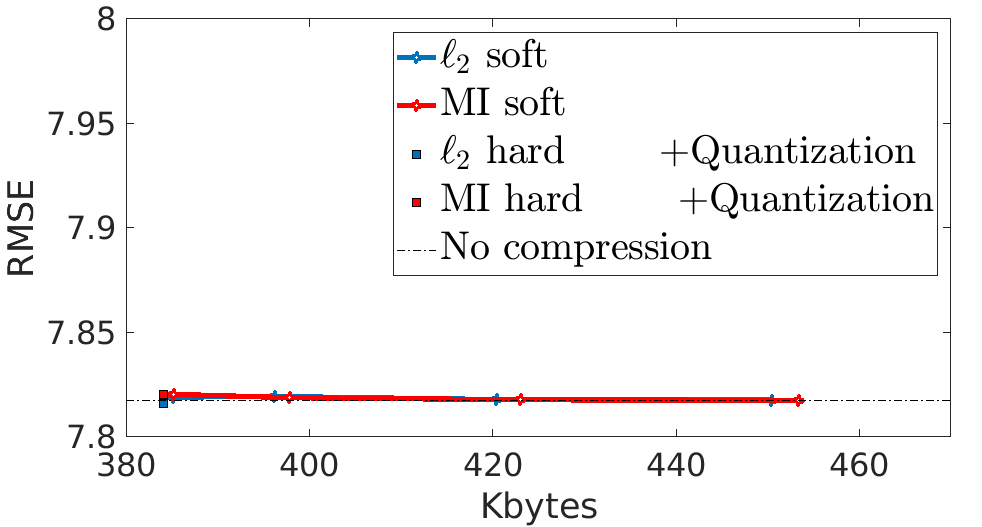}
\includegraphics[width=5.5cm]{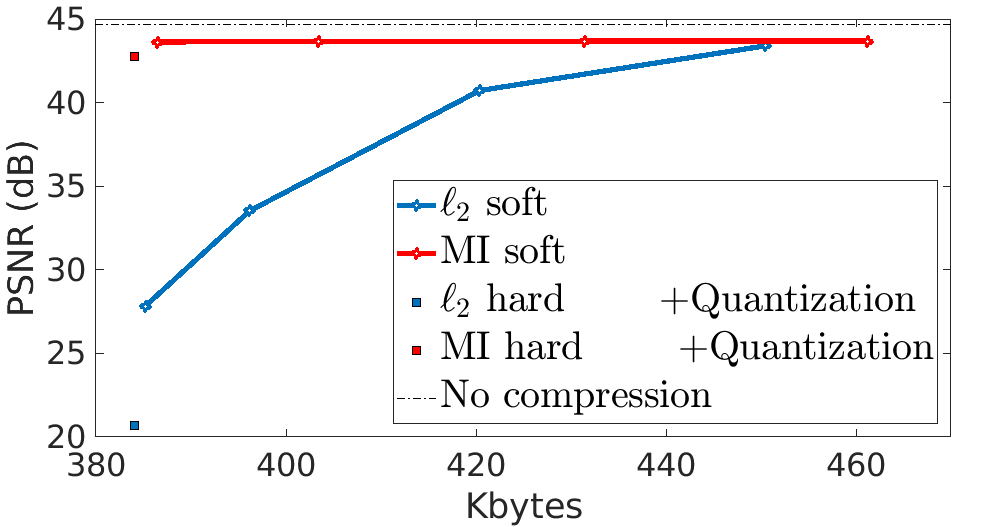} \\
\includegraphics[width=5.5cm]{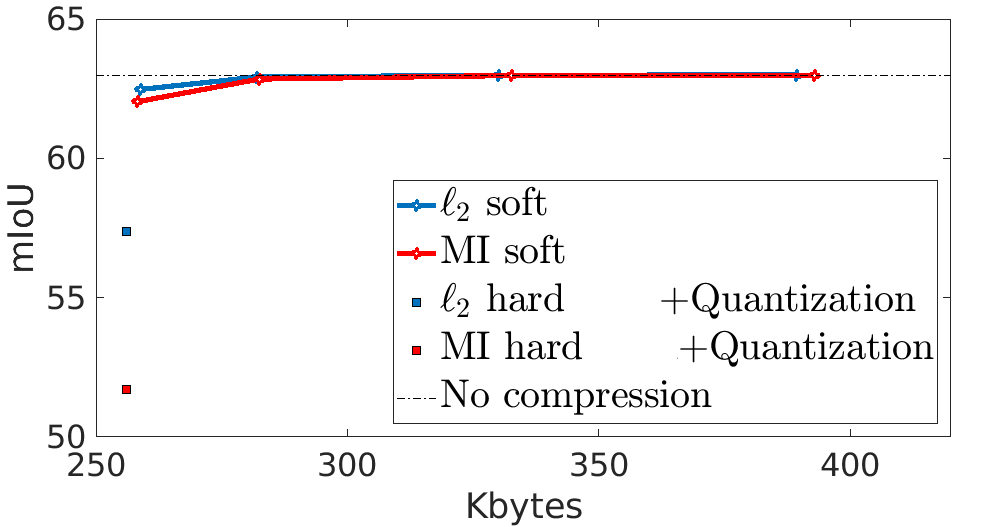}
\includegraphics[width=5.5cm]{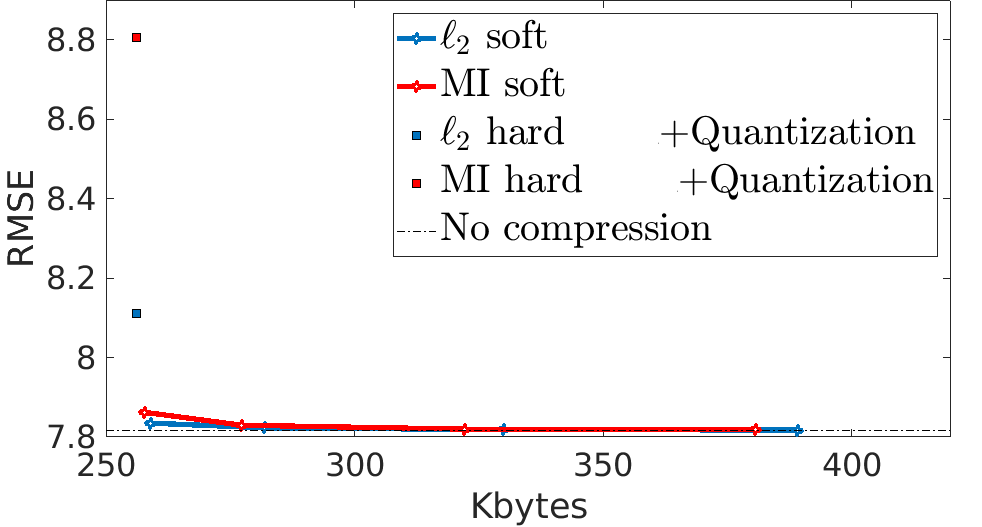}
\includegraphics[width=5.5cm]{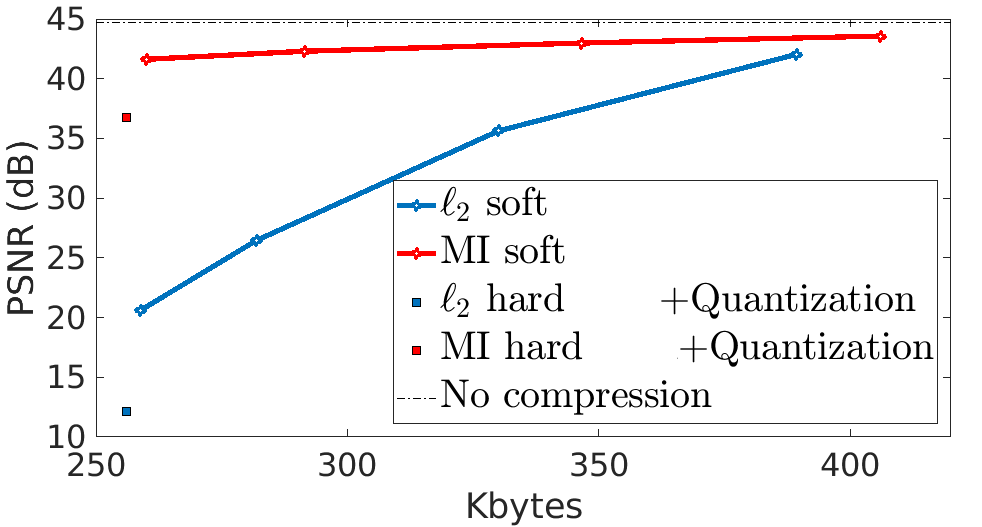}
%\end{tabular}
\caption{Task accuracy vs. file size in soft selection for semantic segmentation (left), disparity map estimation (middle) and input reconstruction (right). Top: $75\%$ base selection. Bottom: $50\%$ base selection.}
\label{fig:multi_soft}
%\vspace{-10pt}
\end{figure*}

\textbf{Soft selection.}
Here, features are sorted according to  importance ($\ell_2$ or MI-based) and 8-bit quantized.  
Then a certain percentage of the most important features is selected as base features. The remaining (enhancement) features are tiled into an image and  compressed using
Kvazaar,\footnote{https://github.com/ultravideo/kvazaar} an open-source HEVC encoder, using four QP values: \{10, 20, 30, 40\}. Lower QP values result in more accurate feature reconstruction at the cost of higher bitrate. 
Fig.~\ref{fig:multi_soft} shows the results in terms of task accuracy vs. average file size in Kbytes, for two cases -- $75\%$ selected as base (first row) and $50\%$ selected as base (second row). The two single dots in each plot show the accuracy with the corresponding hard selection, where the given percentage of features is selected and 8-bit quantized, while others are discarded. We see that soft selection can close the gap between hard selection and default model performance at the cost of slightly increased file size. Also, with $50\%$ base selection, the same task accuracies are achieved with fewer bits compared to $75\%$ base, due to the ability to exercise compression over more enhancement features.

\textbf{Multi-objective analysis.} So far, we looked at the cases where features are selected for one of several tasks. But what if all tasks are needed? This leads to a multi-objective problem. One way to take task preference into account when selecting features in such a scenario is to define task distortion $D_j(C')$ of task $j$ as a fraction of accuracy reduction due to selecting $C'$ features:
\begin{equation}
\label{eq:dist_def}
    D_j(C') = \frac{|A_j(C) - A_j(C')|}{A_j(C)},
\end{equation}
where $A_j(C)$ is the task accuracy with all $C$ features and $A_j(C')$ is the task accuracy with $C'$ most important features \textit{for that task}. 
The total distortion $D(C')$ with $C'$ retained features for a model with $K$ tasks can be defined as: 
\begin{equation}
\label{eq:total_dist}
D(C') = \sum_{j=1}^{K} w_jD_j(C'),
\end{equation}
where the weights $w_j$ are non-negative, sum up to $1$, and reflect task preferences. For a given $K$-tuple of task preference weights $(w_1,w_2,...,w_K)$, one can examine the total distortions~(\ref{eq:total_dist}) produced by various task-specific importance orderings $\mathcal{O}_j$ 
that leads to the lowest total distortion~(\ref{eq:total_dist}).  
In the 3-task model studied here,  $w_1+w_2+w_3=1$ with $0\leq w_1,w_2,w_3 \leq 1$, which defines a triangle in the $(w_1,w_2,w_3)$ space. 
In the next set of experiments, we examine the regions of this triangle where MI-based feature selection achieves lower total distortion~(\ref{eq:total_dist}) compared to $\ell_2$-based selection.

\begin{figure}[tb!]
\includegraphics[width=4cm]{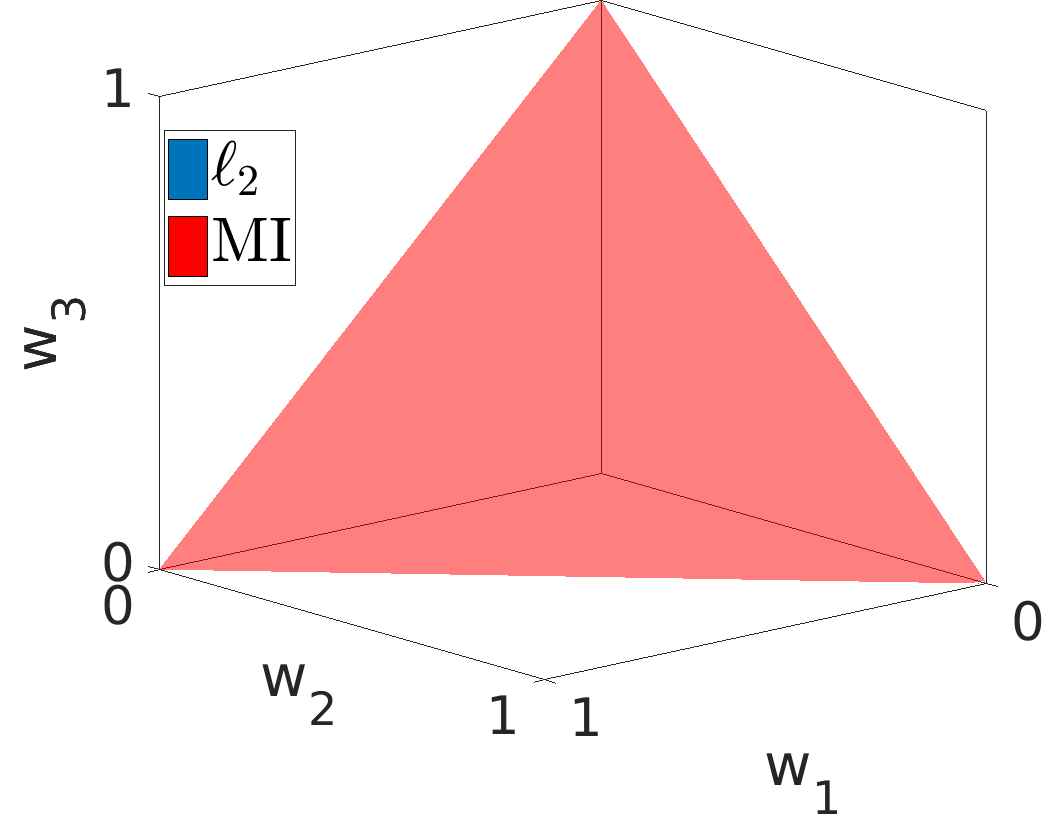} \hspace{5pt}
\includegraphics[width=4cm]{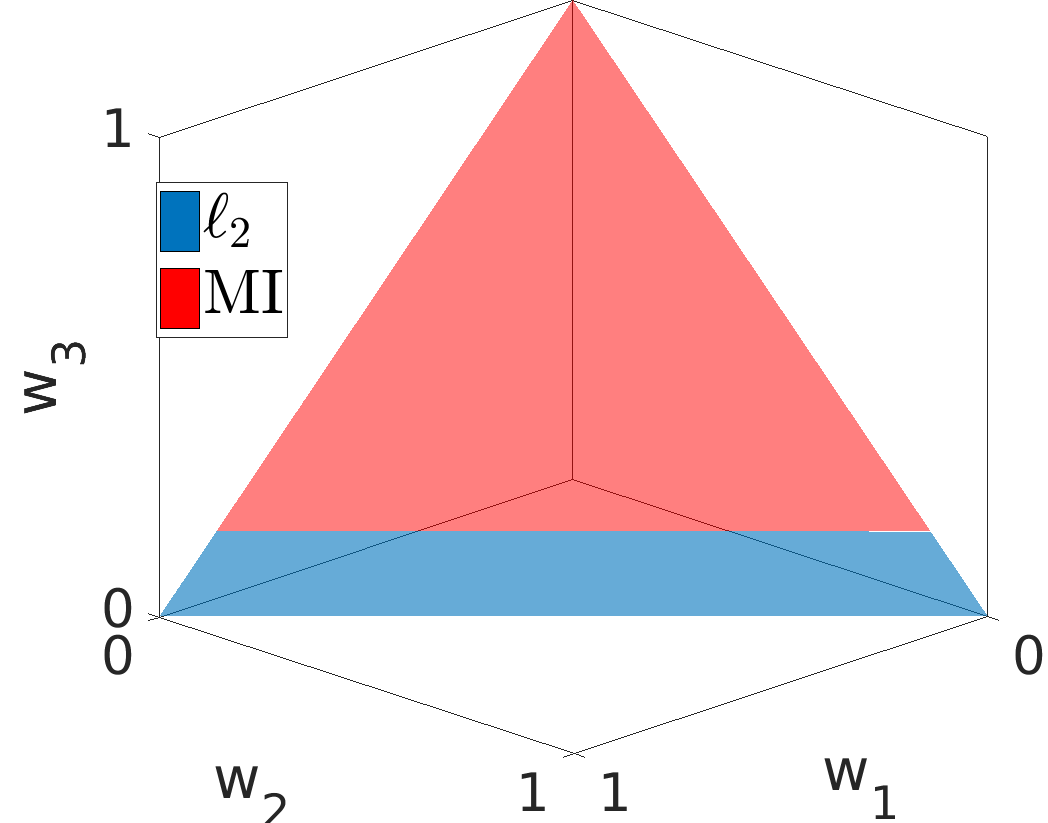} \\ \vspace{5pt} \\
\includegraphics[width=4cm]{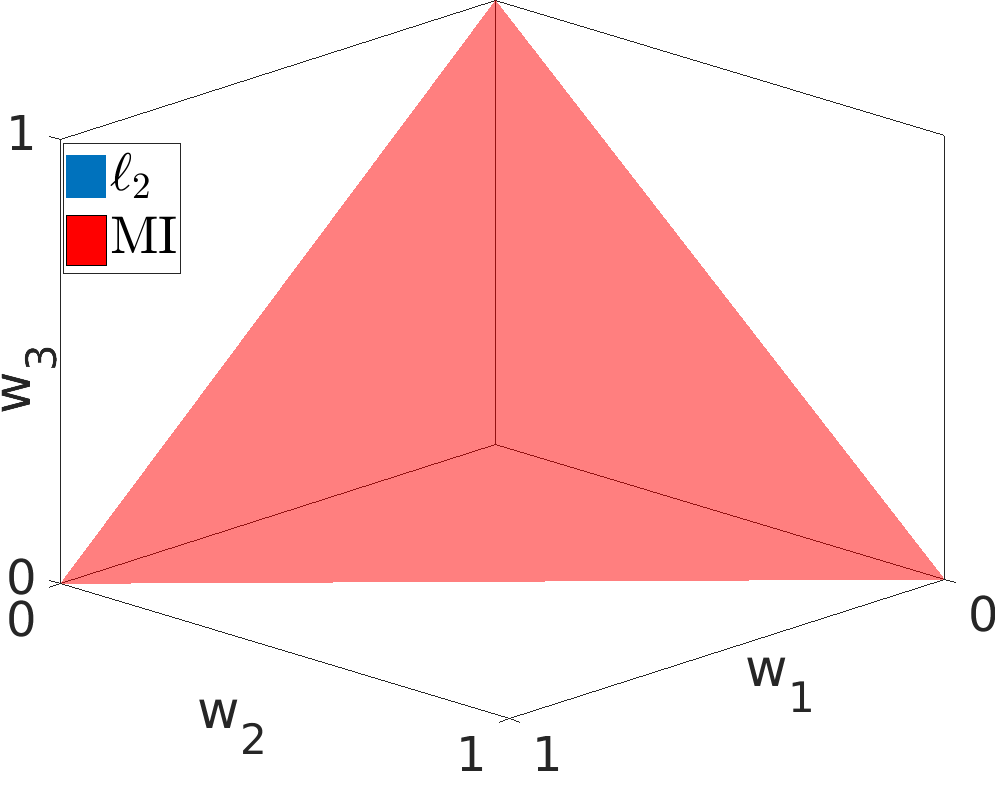} \hspace{5pt}
\includegraphics[width=4cm]{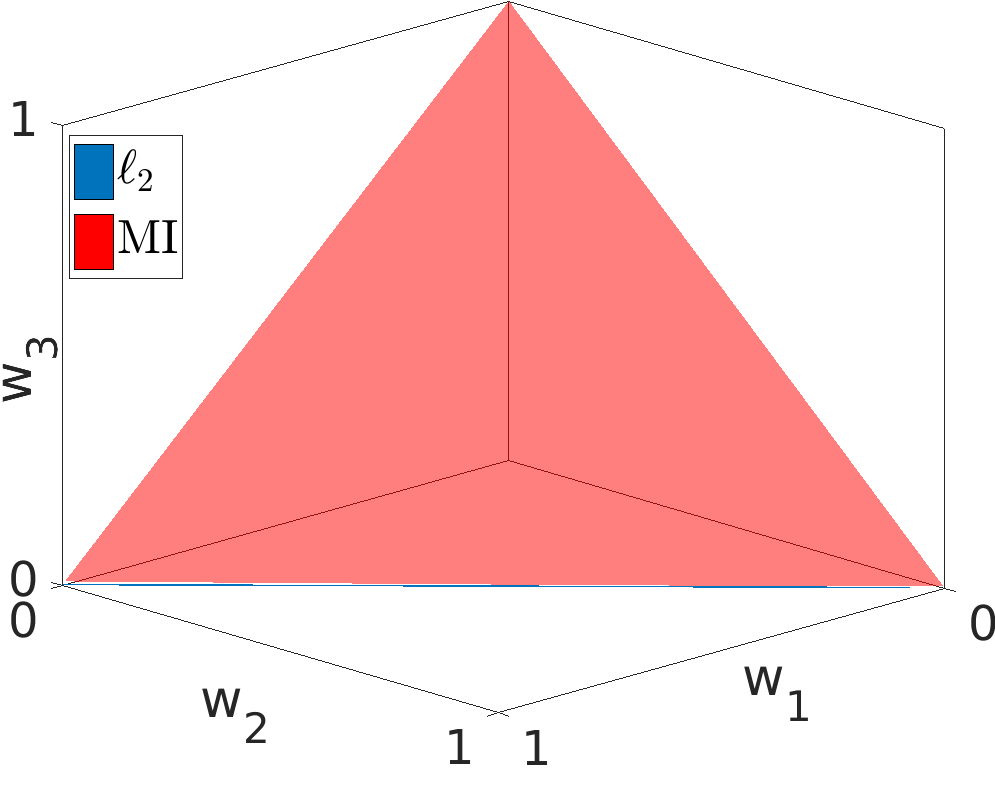}
\caption{The set of task weights $(w_1,w_2,w_3)$ over which a particular feature selection criterion leads to lower total distortion in equation \eqref{eq:total_dist}. Top: hard selection. Bottom: soft selection. Left: $75\%$ selection. Right: $50\%$ selection.}
\label{fig:multi_hard_dist}
%\vspace{-10pt}
\end{figure}

Fig.~\ref{fig:multi_hard_dist} (top) shows which feature selection method -- $\ell_2$ or Mi-based -- leads lower total distortion  at $75\%$ and $50\%$ hard selection.  For each point in the triangle, the total distortion~\eqref{eq:total_dist} is computed for the two selection methods and the point is colored depending on which approach leads to lower total distortion. As seen in the figure, MI-based selection is the better choice across the entire weight space at $75\%$ selection. At $50\%$ selection, $\ell_2$-based approach is preferred near the corners that correspond to semantic segmentation and depth estimation, but so long as input reconstruction holds some importance, the MI-based approach is better. Overall, the MI-based approach wins across 
$74\%$ of the weight space at $50\%$ selection. Fig.~\ref{fig:multi_hard_dist} (bottom) shows the results for $75\%$ and $50\%$ soft selection across all tested QP values. MI-based selection is better across virtually all choices of weights in both cases; the only exception is a small set of weights at the bottom of the triangle for $50\%$ soft selection.

\vspace{-5pt}
\section{Conclusions}
\label{sec:conclusion}
\vspace{-5pt}
We presented hard and soft (compressive) feature selection for multi-task DNNs based on mutual information (MI). Compared to  norm- and proximity-based approaches, the MI-based approach allows one to take task-specific feature importance into account. It was demonstrated that such approach is preferred over a large fraction of the task preference weight space. In addition, we showed that soft selection is able to close the gap between hard selection and default  model performance by using different amounts of feature compression.

\vspace{-5pt}

%\vspace{-0.3cm}
\begin{small}  
\bibliographystyle{IEEEtran}
\bibliography{icme2023template}

\begin{thebibliography}{10}
\providecommand{\url}[1]{#1}
\csname url@samestyle\endcsname
\providecommand{\newblock}{\relax}
\providecommand{\bibinfo}[2]{#2}
\providecommand{\BIBentrySTDinterwordspacing}{\spaceskip=0pt\relax}
\providecommand{\BIBentryALTinterwordstretchfactor}{4}
\providecommand{\BIBentryALTinterwordspacing}{\spaceskip=\fontdimen2\font plus
\BIBentryALTinterwordstretchfactor\fontdimen3\font minus \fontdimen4\font\relax}
\providecommand{\BIBforeignlanguage}[2]{{%
\expandafter\ifx\csname l@#1\endcsname\relax
\typeout{** WARNING: IEEEtran.bst: No hyphenation pattern has been}%
\typeout{** loaded for the language `#1'. Using the pattern for}%
\typeout{** the default language instead.}%
\else
\language=\csname l@#1\endcsname
\fi
#2}}
\providecommand{\BIBdecl}{\relax}
\BIBdecl

\bibitem{edgeAI2022review}
W.~Su, L.~Li, F.~Liu, M.~He, and X.~Liang, ``{AI} on the edge: A comprehensive review,'' \emph{Artif. Intell. Rev.}, vol.~55, no.~8, pp. 6125--6183, Dec. 2022.

\bibitem{neurosurgeon}
Y.~Kang, J.~H.~C. Gao, A.~Rovinski, T.~Mudge, J.~Mars, and L.~Tang, ``Neurosurgeon: Collaborative intelligence between the cloud and mobile edge,'' \emph{SIGARCH Comput. Archit. News}, vol.~45, no.~1, pp. 615--629, Apr. 2017.

\bibitem{shlezinger2022iotm}
N.~Shlezinger and I.~V. Bajić, ``Collaborative inference for {AI}-empowered {IoT} devices,'' \emph{IEEE Internet of Things Magazine}, vol.~5, no.~4, pp. 92--98, 2022.

\bibitem{collab_offloading}
A.~E. Eshratifar and M.~Pedram, ``Energy and performance efficient computation offloading for deep neural networks in a mobile cloud computing environment,'' in \emph{Proc. ACM Great Lakes Symp. on VLSI (GLSVLSI'18)}, 2018, pp. 111--116.

\bibitem{matsubara2022split}
Y.~Matsubara, M.~Levorato, and F.~Restuccia, ``Split computing and early exiting for deep learning applications: Survey and research challenges,'' \emph{ACM Comput. Surv.}, vol.~55, no.~5, pp. 1--30, Dec. 2022.

\bibitem{choi_icip}
H.~Choi and I.~V. Baji\'c, ``Deep feature compression for collaborative object detection,'' in \emph{Proc. IEEE ICIP}, 2018.

\bibitem{intelligent_sensing}
Z.~{Chen}, K.~{Fan}, S.~{Wang}, L.~{Duan}, W.~{Lin}, and A.~C. {Kot}, ``Toward intelligent sensing: Intermediate deep feature compression,'' \emph{IEEE Trans. Image Processing}, vol.~29, 2020.

\bibitem{MPEG-VCM_DCfE}
ISO/IEC, ``Draft call for evidence for video coding for machines,'' ISO/IEC JTC1/SC29/WG11/w19508, Jul. 2020.

\bibitem{MPEG-FCM_CTTC}
------, ``Common test and training conditions for {FCM},'' ISO/IEC JTC1/SC29/WG04/N0427, Oct. 2023.

\bibitem{jpegai2023mmmag}
J.~Ascenso, E.~Alshina, and T.~Ebrahimi, ``The {JPEG AI} standard: Providing efficient human and machine visual data consumption,'' \emph{IEEE MultiMedia}, vol.~30, no.~1, 2023.

\bibitem{liu-etal-2019-multi}
X.~Liu, P.~He, W.~Chen, and J.~Gao, ``Multi-task deep neural networks for natural language understanding,'' in \emph{Proc. ACL}, Jul. 2019, pp. 4487--4496.

\bibitem{crawshaw2020multi-survey}
M.~Crawshaw, ``Multi-task learning with deep neural networks: A survey,'' \emph{arXiv preprint arXiv: arXiv:2009.09796}, 2020.

\bibitem{eva2023cvpr}
Y.~Fang, W.~Wang, B.~Xie, Q.~Sun, L.~Wu, X.~Wang, T.~Huang, X.~Wang, and Y.~Cao, ``{EVA}: Exploring the limits of masked visual representation learning at scale,'' in \emph{CVPR}, 2023, pp. 19\,358--19\,369.

\bibitem{Cover}
T.~M. Cover and J.~A. Thomas, \emph{Elements of Information Theory}, 2nd~ed.\hskip 1em plus 0.5em minus 0.4em\relax Wiley, 2006.

\bibitem{Feature_selection_JMLR_2003}
I.~Guyon and A.~Elisseeff, ``An introduction to variable and feature selection,'' \emph{JMLR}, 2003.

\bibitem{Feature_MI_NCA_2013}
J.~R. Vergara and P.~A. Estévez, ``A review of feature selection methods based on mutual information,'' \emph{Neural Comput. \& Appl.}, 2013.

\bibitem{MIFS-ND_2014}
N.~Hoque, D.~K. Bhattacharyya, and J.~K. Kalita, ``{MIFS-ND:} a mutual information-based feature selection method,'' \emph{Expert Syst. Appl.}, vol.~41, no.~14, pp. 6371--6385, 2014.

\bibitem{beraha2019feature}
M.~Beraha, A.~M. Metelli, M.~Papini, A.~Tirinzoni, and M.~Restelli, ``Feature selection via mutual information: New theoretical insights,'' in \emph{IJCNN}, 2019.

\bibitem{mint}
M.~R. Ganesh, J.~J. Corso, and S.~Y. Sekeh, ``{MINT:} deep network compression via mutual information-based neuron trimming,'' \emph{arXiv preprint arXiv:2003.08472}, 2020.

\bibitem{gradient_MI}
M.~K. Lee, S.~Lee, S.~H. Lee, and B.~C. Song, ``Channel pruning via gradient of mutual information for light-weight convolutional neural networks,'' in \emph{Proc. IEEE ICIP}, 2020.

\bibitem{resnet}
K.~He, X.~Zhang, S.~Ren, and J.~Sun, ``Deep residual learning for image recognition,'' in \emph{CVPR'16}, 2016, pp. 770--778.

\bibitem{soft_filter}
Y.~He, G.~Kang, X.~Dong, Y.~Fu, and Y.~Yang, ``Soft filter pruning for accelerating deep convolutional neural networks,'' in \emph{Proc. IJCAI}, 2018, pp. 2234--2240.

\bibitem{Efficient_ConvNets_ICLR_2017}
H.~Li, A.~Kadav, I.~Durdanovic, H.~Samet, and H.~P. Graf, ``Pruning filters for efficient convnets,'' in \emph{ICLR}, 2017.

\bibitem{scale_importance}
Z.~Liu, J.~Li, Z.~Shen, G.~Huang, S.~Yan, and C.~Zhang, ``Learning efficient convolutional networks through network slimming,'' in \emph{ICCV}, 2017, pp. 2736--2744.

\bibitem{prun_multi_ss}
X.~Chen, Y.~Wang, Y.~Zhang, P.~Du, C.~Xu, and C.~Xu, ``Multi-task pruning for semantic segmentation networks,'' \emph{arXiv preprint arXiv:2007.08386}, 2020.

\bibitem{GM}
Y.~He, P.~Liu, Z.~Wang, Z.~Hu, and Y.~Yang, ``Filter pruning via geometric median for deep convolutional neural networks acceleration,'' in \emph{CVPR}, 2019, pp. 4340--4349.

\bibitem{correlation_filters}
P.~Singh, V.~K. Verma, P.~Rai, and V.~P. Namboodiri, ``Leveraging filter correlations for deep model compression,'' in \emph{Proc. IEEE WACV}, 2020, pp. 824--833.

\bibitem{spectral_prun}
H.~Zhuo, X.~Qian, Y.~Fu, H.~Yang, and X.~Xue, ``Scsp: Spectral clustering filter pruning with soft self-adaption manners,'' \emph{arXiv preprint arXiv:1806.05320}, 2018.

\bibitem{channel_ieee_access}
A.~{Polyak} and L.~{Wolf}, ``Channel-level acceleration of deep face representations,'' \emph{IEEE Access}, vol.~3, 2015.

\bibitem{prune_stat_guide}
H.~Li, C.~Ma, W.~Xu, and X.~Liu, ``Feature statistics guided efficient filter pruning,'' \emph{arXiv preprint arXiv:2005.12193}, 2020.

\bibitem{thinet}
J.~Luo, J.~Wu, and W.~Lin, ``Thinet: A filter level pruning method for deep neural network compression,'' in \emph{ICCV}, 2017.

\bibitem{HRank}
M.~Lin, R.~Ji, Y.~Wang, Y.~Zhang, B.~Zhang, Y.~Tian, and L.~Shao, ``{HRank:} filter pruning using high-rank feature map,'' in \emph{CVPR}, 2020, pp. 1526--1535.

\bibitem{attention_scale}
K.~Yamamoto and K.~Maeno, ``Pcas: Pruning channels with attention statistics for deep network compression,'' \emph{arXiv preprint arXiv:1806.05382}, 2018.

\bibitem{paninski2003mi}
L.~Paninski, ``Estimation of entropy and mutual information,'' \emph{Neural Computation}, vol.~15, no.~6, pp. 1191--1253, 2003.

\bibitem{belghazi2021mine}
M.~I. Belghazi, A.~Baratin, S.~Rajeswar, S.~Ozair, Y.~Bengio, A.~Courville, and R.~D. Hjelm, ``{MINE}: Mutual information neural estimation,'' \emph{PMLR}, vol.~80, pp. 531--540, 2018.

\bibitem{mcallester20aistats}
D.~McAllester and K.~Stratos, ``Formal limitations on the measurement of mutual information,'' in \emph{Proc. Int. Conf. Artificial Intelligence and Statistics (AISTATS)}, Aug. 2020, pp. 875--884.

\bibitem{bellman1957dynamic}
R.~Bellman, \emph{Dynamic Programming}.\hskip 1em plus 0.5em minus 0.4em\relax Princeton University Press, 1957.

\bibitem{saxe}
A.~M. Saxe, Y.~Bansal, J.~Dapello, M.~Advani, A.~Kolchinsky, B.~D. Tracey, and D.~D. Cox, ``On the information bottleneck theory of deep learning,'' in \emph{ICLR}, 2018.

\bibitem{MI_formula}
J.~Boets, K.~D. Cock, and B.~D. Moor, ``A mutual information based distance for multivariate gaussian processes,'' in \emph{Modeling, Estimation and Control}.\hskip 1em plus 0.5em minus 0.4em\relax Springer, 2007, pp. 15--33.

\bibitem{kmeans}
A.~Likas, N.~Vlassis, and J.~J. Verbeek, ``The global k-means clustering algorithm,'' \emph{Pattern Recognition}, vol.~36, no.~2, pp. 451--461, 2003.

\bibitem{Saeed_ICASSP_2020}
S.~R. {Alvar} and I.~V. {Bajić}, ``Bit allocation for multi-task collaborative intelligence,'' in \emph{Proc. IEEE ICASSP}, 2020.

\bibitem{FC}
E.~{Shelhamer}, J.~{Long}, and T.~{Darrell}, ``Fully convolutional networks for semantic segmentation,'' \emph{IEEE Trans. Pattern Anal. Mach. Intell.}, vol.~39, no.~4, pp. 640--651, April 2017.

\bibitem{Cordts2016Cityscapes}
M.~Cordts, M.~Omran, S.~Ramos, T.~Rehfeld, M.~Enzweiler, R.~Benenson, U.~Franke, S.~Roth, and B.~Schiele, ``The {Cityscapes} dataset for semantic urban scene understanding,'' in \emph{CVPR}, 2016.

\bibitem{YOLOv3}
J.~Redmon and A.~Farhadi, ``{YOLOv3}: An incremental improvement,'' \emph{arXiv preprint arXiv:1804.02767}, 2018.

\end{thebibliography}
\end{small}

\end{document}